\documentclass[aps,prb,amsmath,amssymb,reprint,superscriptaddress]{revtex4-2}

\usepackage{graphicx}
\usepackage{amsmath,amssymb}
\usepackage{xcolor}
\usepackage{color,soul}

\begin{document}

\title{Structural, Bonding, and Optical Properties of B$_{18}$Ca$_2$ Clusters: Double-Ring Forms}

\author{P. L. Rodr\'iguez-Kessler}
\email{plkessler@cio.mx}
\affiliation{Centro de Investigaciones en \'Optica A.C., Loma del Bosque 115, Lomas del Campestre, Leon, 37150, Guanajuato, Mexico}

\date{\today}

\begin{abstract}
The structural and electronic properties of the doubly calcium-doped boron cluster B$_{18}$Ca$_2$ have been systematically investigated using density functional theory calculations. Basin-hopping searches reveal that B$_{18}$Ca$_2$ adopts a double-ring geometry as its global minimum, consisting of two fused B$_9$ rings symmetrically stabilized by calcium atoms located above and below the boron framework. Vibrational frequency calculations verify the dynamical stability of the low-lying structures, while infrared and UV-Vis spectra highlight strong Ca--B coupling and pronounced electronic delocalization within the boron scaffold. Atomic dipole-corrected Hirshfeld charge analysis indicates substantial charge transfer from Ca to the electron-deficient boron framework, with the donated electrons uniformly delocalized over the B$_{18}$ skeleton. Real-space bonding analyses based on the electron localization function (ELF), Interaction Region Indicator (IRI), and the Laplacian of the electron density reveal an extended multicenter bonding network characterized by global $\sigma$-delocalization and Ca-induced polarization effects rather than localized two-center Ca--B bonds. Together, these results establish B$_{18}$Ca$_2$ as a prototypical boron toroidal cluster and provide fundamental insights into the role of alkaline-earth doping in stabilizing complex boron nanostructures.
\end{abstract}

\maketitle

\section{Introduction}

Boron clusters continue to attract significant attention in modern cluster science due to their structural versatility and the rich landscape of electron-deficient, multicenter bonding they exhibit.\cite{C6CC09570D,C9CP03496J} Among them, medium-sized species such as the bare B$_{18}$ cluster are known to favor double-ring or toroidal geometries, providing highly delocalized frameworks capable of supporting complex multicenter bonding patterns.\cite{doi:10.1021/jp8087918,C6SC02623K} The incorporation of metal atoms into boron clusters has unlocked a variety of unprecedented structural and electronic motifs, including inverse-sandwich complexes,\cite{JIA2014128,Zhuan-Yu2014,PHAM2019186,C5CP01650A,LI202325821,RODRIGUEZKESSLER2025117486,b7al2,b7cr2} metal-centered aromaticity, and fully delocalized toroidal frameworks.\cite{doi:10.1021/acs.inorgchem.7b02585,doi:10.1021/acs.jpclett.0c02656} In this context, doubly doped systems of the form B$_{18}$M$_2$ (M = metal) represent a largely underexplored chemical space, where the extended boron scaffold can stabilize two metal centers through symmetric multicenter interactions.

Calcium doping is particularly intriguing. As an alkaline-earth element with relatively diffuse 4s and 3d-like valence functions, Ca readily donates electrons to electron-deficient boron frameworks, promoting the formation of highly connected Ca–B bonding networks. This behavior parallels trends observed in smaller inverse-sandwich clusters such as B$_7$M$_2$, where two metal centers cap a single boron ring and contribute to extensive $\sigma$- and $\pi$-delocalization.\cite{D5CP01078K,b8cu3,B7Y2,GUEVARAVELA2025115487}

In this work, we present a comprehensive computational characterization of the B$_{18}$Ca$_2$ cluster, combining global optimization, DFT energetics, vibrational analysis, charge-, electron density-based analyses, and AdNDP. Our results reveal that B$_{18}$Ca$_2$ adopts a robust double-ring geometry that effectively stabilizes two calcium centers through symmetric multicenter Ca–B interactions, establishing this species as a new prototype of metal-stabilized boron toroids.\cite{C7CP04158F} Moreover, this work serves to expand the libraries of doped boron clusters for potential applications.\cite{OLALDELOPEZ2024}

\section{Computational Details}

Calculations performed in this work are carried out by using density functional theory (DFT) as implemented in the ORCA 6.0.0 code.\cite{10.1063/5.0004608} The exchange and correlation energies are addressed by the PBE0 functional in conjunction with the Def2-TZVP basis set.\cite{10.1063/1.478522,B508541A} Atomic positions are self-consistently relaxed through a Quasi-Newton method employing the BFGS algorithm. The SCF convergence criterion is set to TightSCF in the input file. This results in geometry optimization settings of 1.0e$^{-08}$ Eh for total energy change and 2.5e$^{-11}$ Eh for the one-electron integrals. The  Van  der  Waals  interactions  are  included in the exchange-correlation functionals with empirical dispersion corrections of Grimme DFT-D3(BJ). The electron localization function (ELF) was computed and analyzed using Multiwfn.\cite{https://doi.org/10.1002/jcc.22885} Global minima searches were conducted using standard basin-hopping (BH) algorithm with random rotational–translational perturbations and subsequent DFT local optimization at the PBE0/def2-SVP level.\cite{basin} Low-lying candidates ($<$20 kcal/mol) were reoptimized at PBE0/def2-TZVP level. Vibrational frequency calculations confirmed all minima as true stationary points (no imaginary modes). Spin states from singlet to quintet were evaluated. Spatial region analyses (calc. grid data) were calculated with Multiwfn.

\section{Results and Discussion}

\begin{figure}[ht]
\centering
\includegraphics[width=0.45\textwidth]{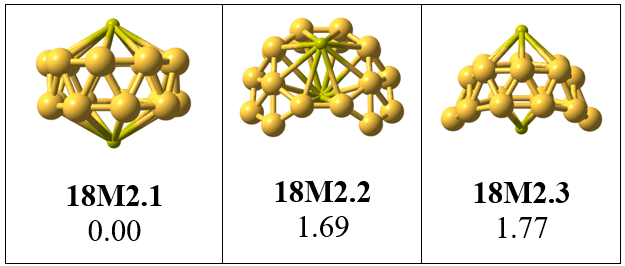}
\caption{Lowest energy structures for B$_{18}$Ca$_2$ cluster at PBE0/def2-TZVP.}
\label{fig_geom}
\end{figure}

The low-energy structures of the B$_{18}$Ca$_2$ cluster exhibit a pronounced preference for compact and symmetric motifs dominated by a double-ring boron framework stabilized by calcium doping (see Fig.~\ref{fig_geom}). The global minimum adopts a D$_{3h}$-like geometry, in which the B$_{18}$ skeleton is composed of two fused B$_9$ rings. In this configuration, the Ca atoms are located above and below the center of the boron rings, maximizing electrostatic stabilization through charge transfer within the boron network. The higher-lying isomers retain the same fundamental double-ring topology but exhibit subtle distortions in ring planarity and variations in the Ca coordination environments, resulting in modest energetic penalties. Of course a number of isomers with intermediate structures between these isomers results from the structural search, but we show the most representative ones. Overall, the structural evolution of B$_{18}$Ca$_2$ reflects a balance between multicenter boron–boron bonding and predominantly ionic boron–calcium interactions, giving rise to a well-defined family of energetically competitive double-ring configurations.

\subsection{Relative energies}

The relative energies clearly indicate a strong preference for the double-ring structure, which is identified as the global minimum at both the PBE0 and $\omega$B97X-D3 levels of theory, underscoring its superior structural stability. In contrast, the endohedral and bent double-ring isomers are significantly less stable, lying higher in energy by approximately 1.6–1.8~eV, which suggests that they are thermodynamically inaccessible under equilibrium conditions. Notably, the excellent agreement between PBE0 and $\omega$B97X-D3 confirms that the isomer energy ordering is insensitive to the choice of exchange–correlation functional, thereby reinforcing the double-ring motif as the dominant structural feature for this cluster composition.

\begin{table}[h]
\centering
\setlength{\tabcolsep}{12pt} 
\caption{Relative energies of the lowest B$_{18}$Ca$_2$ isomers (in eV). The global minimum is denoted by GM.}
\begin{tabular}{lcc}
\hline\hline
Isomer & PBE0 & $\omega$B97X-D3 \\
\hline
Double-ring structure (GM) & 0.0 & 0.0 \\
Endohedral structure & 1.69 & 1.61 \\
Bent double-ring isomer & 1.77 & 1.73 \\
\hline\hline
\end{tabular}
\end{table}

\subsection{Vibrational and Optical Properties}

The vibrational and electronic absorption properties of the B$_{18}$Ca$_2$ double-ring cluster were investigated to elucidate the interplay between structure, bonding, and spectroscopic response. The calculated infrared (IR) spectrum is dominated by low- and mid-frequency vibrational modes, reflecting the collective dynamics of the boron framework and its stabilization by calcium atoms. Strong IR-active modes appearing below approximately 450~cm$^{-1}$ are primarily associated with Ca--B stretching and Ca--B--B bending motions. The most intense absorptions, located around 410--415~cm$^{-1}$, indicate pronounced charge displacement along the Ca--B interactions, highlighting the electrostatic coupling between the calcium centers and the electron-deficient boron rings. These modes provide direct vibrational evidence of the stabilizing role played by Ca atoms in the double-ring architecture. In the intermediate frequency range (500--800~cm$^{-1}$), several moderately intense bands are observed and can be assigned to in-plane and out-of-plane B--B skeletal vibrations of the double-ring motif. The presence of closely spaced peaks in this region suggests partial symmetry breaking induced by calcium coordination, which lifts vibrational degeneracies characteristic of pristine boron clusters. Higher-frequency modes above 900~cm$^{-1}$ exhibit comparatively low intensities and correspond to more localized B--B stretching vibrations, consistent with the absence of terminal bonds in the cluster.

\begin{table*}[ht]
\centering
\begin{tabular}{cc}
 \resizebox*{0.40\textwidth}{!}{\includegraphics{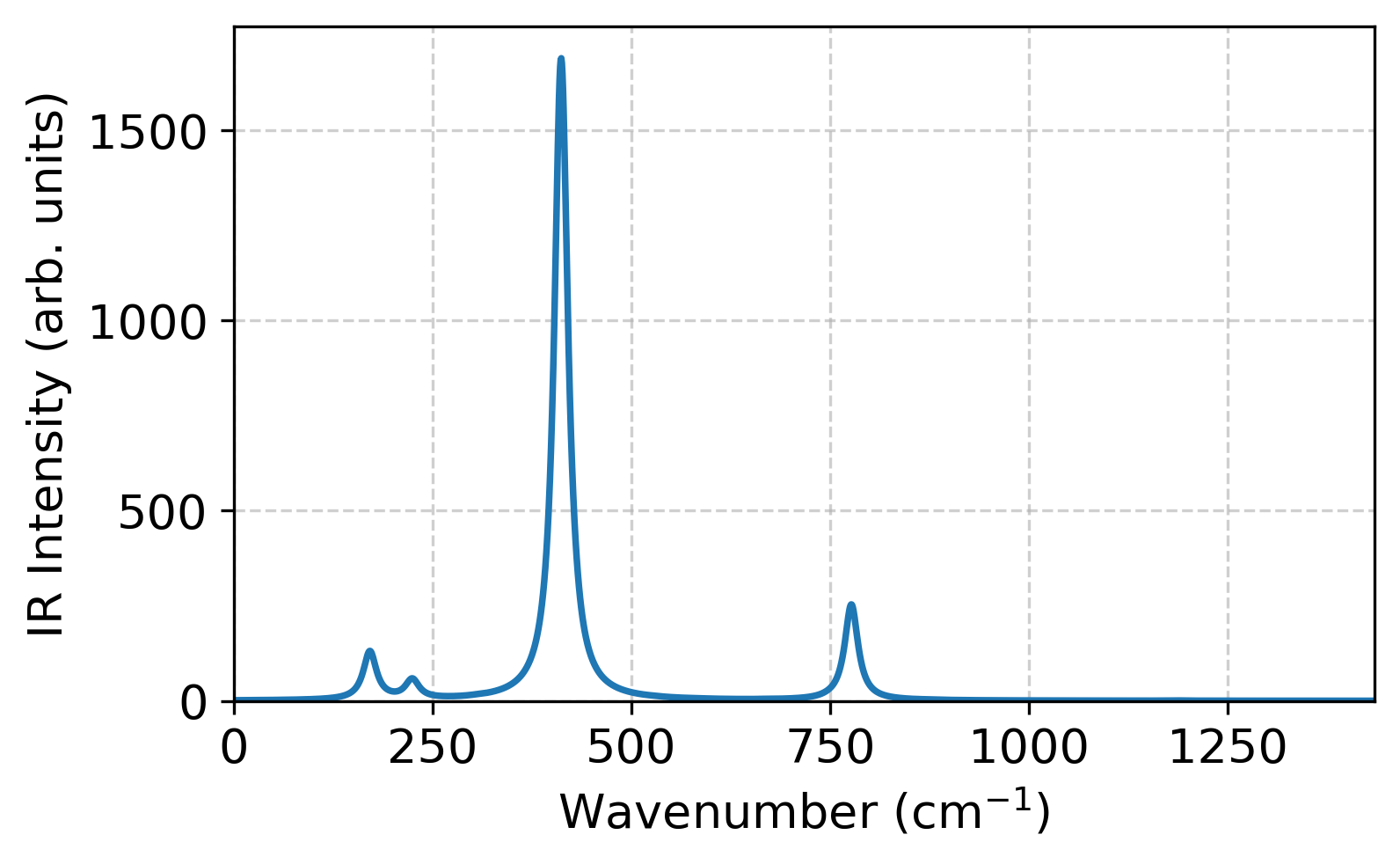}} &
\resizebox*{0.39\textwidth}{!}{\includegraphics{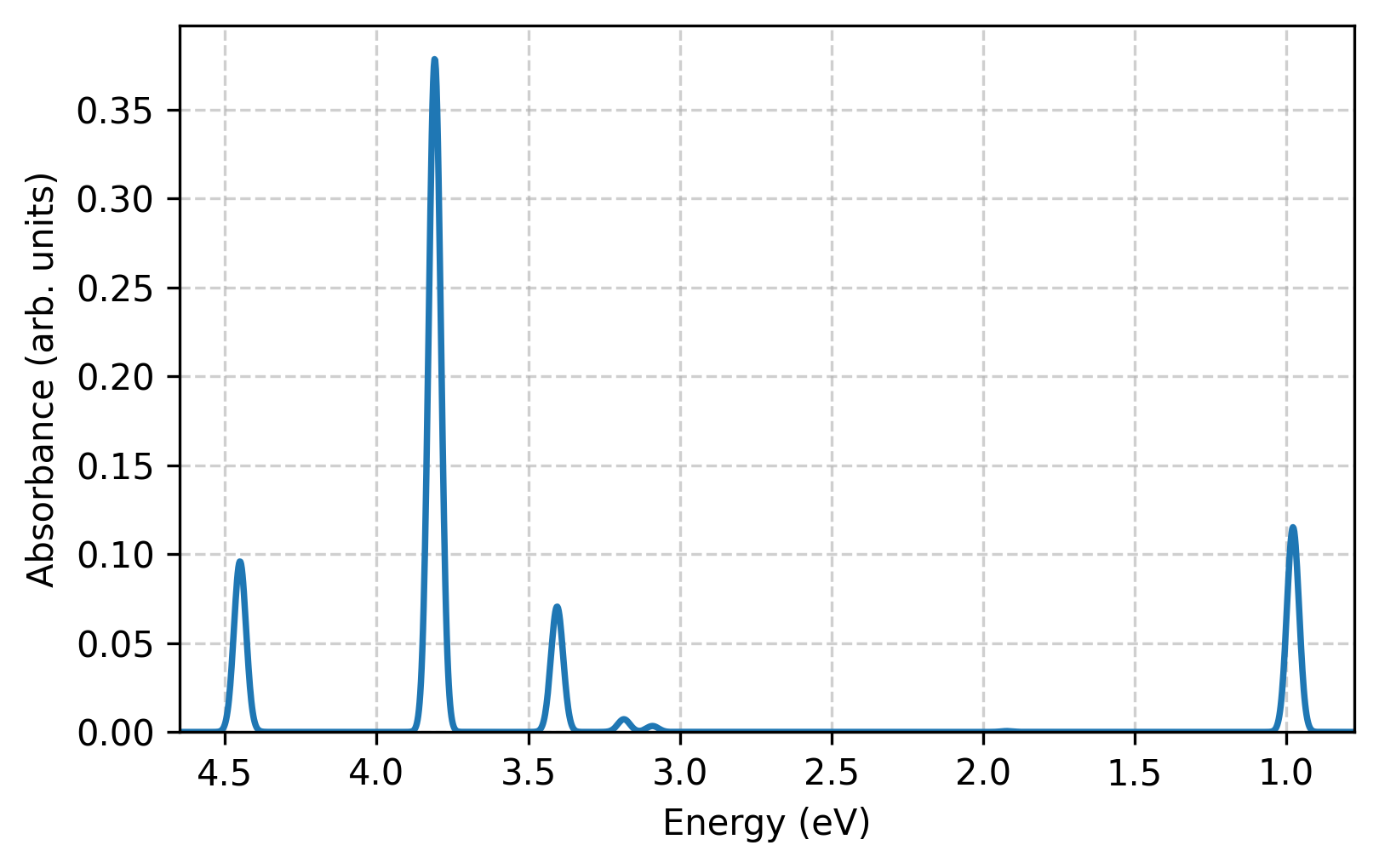}} \\
\end{tabular}
\caption{IR and UV-VIS spectra of B$_{18}$Ca$_2$ cluster.}
\label{fig_elf_lol}
\end{table*}

The UV--Vis absorption spectrum of B$_{18}$Ca$_2$ spans from the near-infrared to the near-ultraviolet region, revealing a rich electronic excitation manifold. Low-energy absorptions around 1.0~eV are characterized by significant oscillator strengths and are attributed to charge-transfer excitations involving calcium-derived states and the delocalized boron framework. These transitions reflect efficient electronic communication between the metal centers and the boron rings. More intense absorption features appear in the near-UV region, particularly around 3.7--3.9~eV, where strong $\pi \rightarrow \pi^*$ transitions delocalized over the double-ring boron skeleton dominate the spectrum. The large oscillator strengths and partial degeneracy of these transitions underscore the high degree of electronic delocalization and the role of calcium atoms as electron donors that enhance the optical activity of the cluster. Overall, the combined vibrational and optical signatures confirm that the B$_{18}$Ca$_2$ double-ring cluster is both vibrationally robust and optically active, with properties governed by the synergistic interaction between the boron framework and calcium dopants.

\subsection{Charge analysis}

To obtain a more realistic description of charge transfer in the B$_{18}$Ca$_2$ cluster, atomic dipole–corrected Hirshfeld (ADCH) charges were evaluated. In contrast to the conventional Hirshfeld analysis, the ADCH method accounts for intra-atomic polarization effects and therefore provides a more reliable estimate of charge separation in systems with delocalized bonding. The ADCH results reveal a substantial electron transfer from the calcium dopants to the boron framework, with each Ca atom carrying a pronounced positive charge of approximately +0.96 $e^-$. Correspondingly, the excess electronic density is evenly distributed over the eighteen boron atoms, which each acquire a modest negative charge in the range of -0.10 to -0.11 $e^-$. The near-uniformity of the boron charges indicates a highly delocalized electron distribution within the double-ring B$_{18}$ scaffold, consistent with multicenter boron–boron bonding. The almost identical charges on the two Ca atoms further reflect their symmetric placement above and below the boron rings and confirm their equivalent electronic roles. Overall, the ADCH charge analysis supports a bonding picture in which calcium acts as an efficient electron donor, stabilizing the electron-deficient boron framework through predominantly ionic B–Ca interactions, while the donated electrons are extensively delocalized over the boron rings. This enhanced charge transfer, compared to standard Hirshfeld values, highlights the critical role of calcium-induced polarization in reinforcing the stability of the B$_{18}$Ca$_2$ double-ring motif.





\subsection{Electron Density–Based Bonding Analysis}

The ELF maps of B$_{18}$Ca$_2$ reveal a bonding pattern in which localized Ca-centered basins coexist with highly delocalized electron density along the boron framework. In the side view, each Ca atom exhibits intense ELF maxima consistent with polarized Ca–B interactions, reflecting substantial Ca→B electron donation and the formation of multicenter Ca–B–B bonding domains that anchor the metal atoms to both boron rings. In contrast, the top-view ELF slice displays a nearly continuous annular distribution around the 18-member double ring, indicative of an extended $\sigma$-delocalized network rather than localized two-center B–B bonds. The uniform circular ELF pattern confirms that the boron scaffold supports multicenter aromatic-like electron circulation, while the low ELF density in the central cavity rules out inner-core bonding. Together, these features demonstrate that B$_{18}$Ca$_2$ is stabilized by the synergistic combination of Ca-induced charge transfer and global $\sigma$-delocalization across the toroidal B$_{18}$ skeleton.

\begin{table*}[ht]
\centering
\begin{tabular}{cc}
 \resizebox*{0.40\textwidth}{!}{\includegraphics{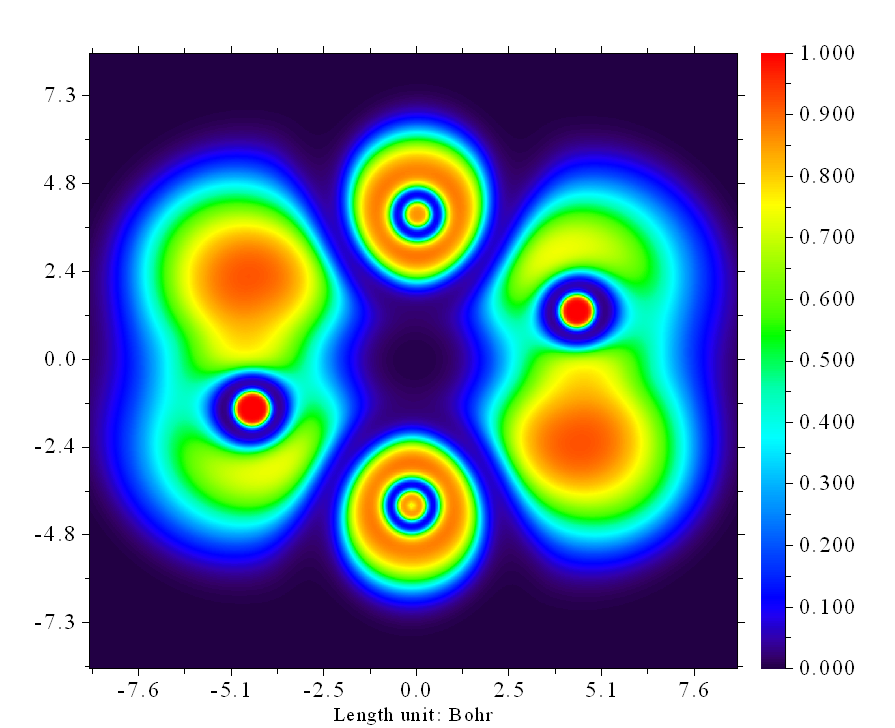}} &
\resizebox*{0.40\textwidth}{!}{\includegraphics{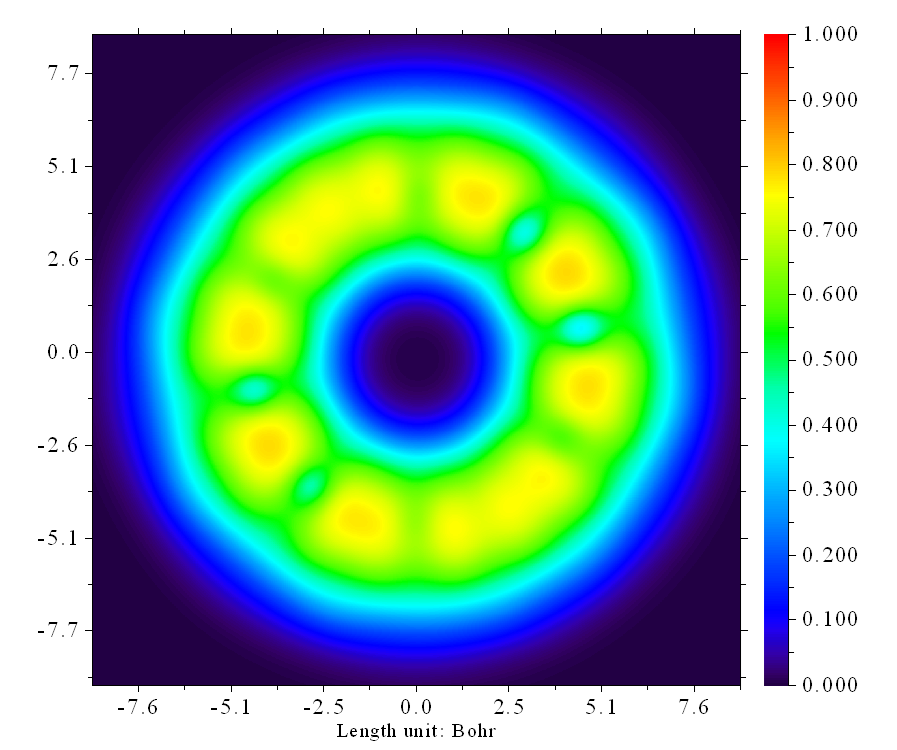}} \\
\end{tabular}
\caption{ELF 2D maps for B$_{18}$Ca$_2$ cluster.}
\label{fig_elf_lol}
\end{table*}


The Laplacian of the electron density ($\nabla^2\rho$) for B$_{18}$Ca$_2$ highlights the characteristic distribution of charge concentration and depletion that underpins its multicenter bonding framework. Regions of negative $\nabla^2\rho$ around the B–B contacts indicate localized charge concentration associated with the multi-centered $\sigma$-bonds within each boron ring, whereas the predominantly positive Laplacian surrounding the Ca centers reflects charge depletion consistent with substantial Ca→B electron donation rather than covalent Ca–B bonding. This pattern corroborates the electron-transfer–driven stabilization mechanism inferred from ELF analysis. Complementarily, the Interaction Region Indicator (IRI) map reveals a continuous green toroidal belt encircling the boron framework, characteristic of extended, noncovalent but strongly attractive delocalized interactions that unify the two rings into a coherent $\sigma$-aromatic manifold. The absence of sharp red or blue IRI features between the Ca atoms and the boron cage further confirms that Ca participates through diffuse, multicenter electrostatic interactions rather than localized two-center bonds. 

\begin{figure}[ht]
\centering
\resizebox*{0.40\textwidth}{!}{\includegraphics{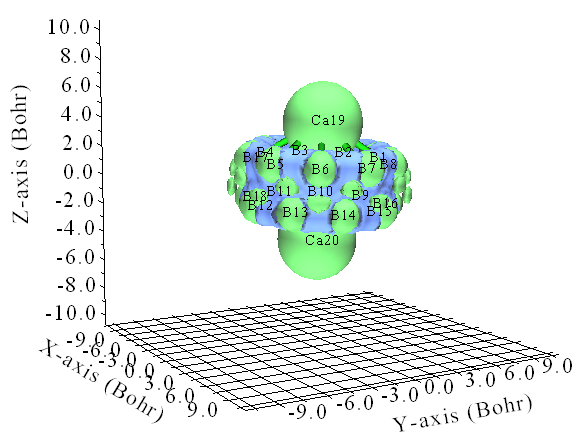}} 
\resizebox*{0.40\textwidth}{!}{\includegraphics{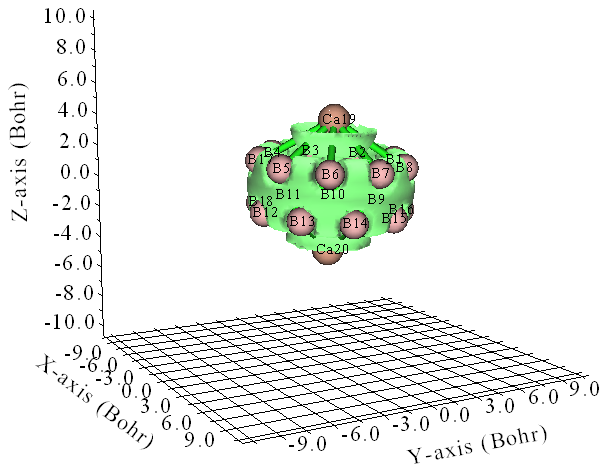}} 
\caption{Laplacian of electron density ($\nabla^2\rho$) and Interaction region indicator (IRI) of B$_{18}$Ca$_2$ clusters.}
\label{fig_elf_lol}
\end{figure}




\section{Conclusions}

In summary, a comprehensive computational investigation has been carried out to elucidate the structural, electronic, and bonding characteristics of the doubly calcium-doped boron cluster B$_{18}$Ca$_2$. Global optimization combined with density functional theory calculations demonstrates that this cluster exhibits a pronounced preference for a compact double-ring geometry, in which two fused B$_9$ rings are symmetrically stabilized by calcium atoms positioned above and below the boron framework. This structural motif is found to be robust across multiple exchange--correlation functionals and represents the global minimum on the potential energy surface. Vibrational frequency and spectroscopic analyses confirm the dynamical stability of the low-energy structures and reveal strong coupling between the calcium centers and the boron scaffold, manifested through characteristic Ca--B vibrational modes and low-energy charge-transfer excitations in the electronic absorption spectrum. Atomic dipole--corrected Hirshfeld charge analysis indicates substantial electron donation from calcium to the electron-deficient boron framework, resulting in a highly delocalized charge distribution over the B$_{18}$ skeleton. Real-space bonding analyses based on the electron localization function, Interaction Region Indicator, and the Laplacian of the electron density consistently show that the stability of B$_{18}$Ca$_2$ arises from a synergistic interplay between Ca-induced polarization effects and extended multicenter $\sigma$-delocalization within the boron double ring, rather than from localized two-center Ca--B bonds. These findings establish B$_{18}$Ca$_2$ as a representative example of a metal-stabilized boron toroidal cluster and highlight the crucial role of alkaline-earth metal doping in engineering stable, delocalized boron nanostructures. The insights gained in this work provide a foundation for the rational design of larger metal-doped boron clusters with tunable structural and electronic properties.


\section{Acknowledgments}
P.L.R.-K. would like to thank the support of CIMAT Supercomputing Laboratories of Guanajuato and Puerto Interior. 



\bibliographystyle{unsrt}
\bibliography{mendelei.bib}
\end{document}